\documentclass[10pt,a4paper]{article}

\usepackage{times}

\usepackage{epsfig,wrapfig}
\usepackage{epstopdf} 
\usepackage[dvipsnames]{xcolor}
\usepackage{fancyhdr}

\usepackage[lmargin=2.5cm, rmargin=2.5cm,tmargin=3.50cm,bmargin=3.50cm]{geometry}
\usepackage{indentfirst}

\usepackage{titlesec}
\titleformat{\section}[hang]
  {\centering}{\thesection}{1ex}{\normalsize \textsc}
\titleformat{\subsection}[hang]
  {}{\thesubsection}{1ex}{\normalsize \textit}

\renewcommand{\thesection}{ \normalsize \textnormal{\Roman{section}.}}
\renewcommand{\thesubsection}{\normalsize \textnormal{\textsc{\textit{\Alph{subsection}.}}}}

\usepackage{graphicx}
\usepackage{amsmath,amssymb}
\usepackage{subcaption}
\usepackage{dcolumn}
\usepackage{bm}
\usepackage{cite}

\def\e{\begin{equation}}
\def\f{\end{equation}}
\def\_#1{{\bf #1}}
\def\@#1{_{\rm #1}}
\def\.{\cdot}

\def\Re{{\rm Re\mit}}

\def\l#1{\label{eq:#1}}
\def\r#1{(\ref{eq:#1})}

\begin{document}

\title{\large \textbf{Non-Scattering Multi-Mirror Systems for Field Localization}}
%
\def\affil#1{\begin{itemize} \item[] #1 \end{itemize}}
\author{\normalsize \bfseries \underline{F.~S.~Cuesta}${\,}^{1}$, V.~S.~Asadchy${\,}^{1,2}$, M.~S.~Mirmoosa${\,}^{1}$, and S.~A.~Tretyakov${\,}^{1}$
\\ }
\date{}
\maketitle
\thispagestyle{fancy} 
\vspace{-6ex}
\affil{\begin{center}\normalsize ${\,}^{1}$Aalto University, Department of Electronics and Nanoengineering\\ P.O.~Box 15500, FI-00076 Aalto, Finland\\
${\,}^{2}$Stanford University, Department of Electrical Engineering, 94305, Stanford, California, USA\\
francisco.cuestasoto@aalto.fi
 \end{center}}

\begin{abstract}
\noindent \normalsize
\textbf{\textit{Abstract} \ \ -- \ \
Resonators composed of two mirrors, such as Fabry-P{\'e}rot cavities, provide a simple but effective approach to achieve wave transmission with high finesse. Further increase of the quality factor requires reflectors with higher conductivity or multi-mirror solutions. However, the analytical complexity of resonators with more than two mirrors prevents the design of optimal structures without recurring to numerical methods. This work considers an alternative approach to this problem by using cavities which do not produce any electromagnetic scattering. We demonstrate that these invisible cavities can be placed inside one another, resulting in a ``matryoshka-doll''-like resonator. The standing-wave distribution of the inner resonator can be controlled  without disturbing that of the outer one, whereas the whole system remains non-scattering.}
\end{abstract}

\section{Introduction}
The Voyager 1, a space probe launched by NASA in the 70'{}s, is the most distant human-made object from the Earth. In order to reach places as far as Saturn, this probe pushed the limits of science and engineering by using the gravity assist provided by Jupiter, boosting its kinetic energy without using its limited propellant. Also in electromagnetics,  the magnitude of waves in a confined volume can be dramatically increased by an ``assisted boost'' without any additional power source.  The structure required for this purpose is an open resonator, such as a Fabry-P{\'e}rot (FP) cavity.

A FP resonator can be formed by two parallel semi-transparent mirrors, usually with the same characteristics, and its finesse is given by the distance between the mirrors and their reflectance~\cite{Fabry1899,Perot1899}. With the proper design, a FP resonator can be transparent (in terms of zero reflection and unitary transmission), while a standing wave is confined between the mirrors~\cite{Clarke1982,Pan2017,Couston2015}. This kind of field localization can be exploited for applications associated with sensing and field modulation. Furthermore, it could be advantageous  if the standing wave of the resonator with already enhanced amplitude  is used to excite a secondary FP resonator, placed \emph{inside} the first one (by the principle of a    ``matryoshka'' doll). However, exploiting conventional  symmetric FP resonators, it is challenging to find a useful and intuitive analytic solution for the required geometry  without solving the scattering produced between the four mirrors~\cite{Stadt1985}. In this scenario, the most effective approach is to use numerical methods to determine out the critical parameters, like the distance between the  mirrors. Nevertheless, all of these issues can be avoided if the inner resonator is designed so that it does not interfere with the   standing wave of the outer resonator. 

In this talk, we compare the characteristics of a ``matryoshka''-like resonator made of two invisible cavity resonators~\cite{Cuesta2018} with a four-mirror FP resonator in terms of design complexity and performance. Due to the properties of the invisible resonator, we   demonstrate that the matryoshka-like resonator remains non-scattering (invisible), regardless of the inner resonator position. This property can be used to engineer the field distribution inside the cavity. As a proof of concept, in this talk we will also discuss the implementation of a one-dimensional optical {\it all-dielectric} cavity which generates no scattering of incident waves in both backward and forward directions.



\section{Field Modulation Using Matryoshka-like Invisible Resonators}

Let us consider the invisible (non-scattering) resonator proposed in Reference~\cite{Cuesta2018} composed by two infinite metasurfaces separated by a distance $d\@{op}$ in a homogeneous medium (for simplicity, this work considers vacuum). This structure becomes invisible   under two conditions: i) The distance between metasurfaces is equal to an integer of half of the operational wavelength $d\@{op}=n\lambda\@{op}/2$, and ii) the grid impedance of one metasurface should be the negative of the other one, i.e. $Z\@{e1}=-Z\@{e2}=Z\@{e}$. 
Under those conditions, the standing wave ratio (SWR) for this resonator depends on the grid impedances as
\e {\rm SWR}_1 = \dfrac{\vert 2 Z\@{e} - \eta\@{0} \vert +\eta\@{0}}{\vert 2 Z\@{e} - \eta\@{0} \vert -\eta\@{0}}, \l{s_e_swr}\f
where $\eta_0$ corresponds to the characteristic impedance of the medium (vacuum in this case). Notice that low grid impedance values of the two sheets (close to that of a  perfect electric conductor) imply high field contrast in the cavity.
\begin{figure}[t!]
\centering
	\begin{subfigure}{0.01\textwidth}
	\captionsetup{labelformat=empty}
		\caption{${}$}
		\label{fig:structure}
	\end{subfigure}
			\qquad
				\begin{subfigure}{0.01\textwidth}
	\captionsetup{labelformat=empty}
		\caption{${}$}
		\label{fig:mat_fields}		
	\end{subfigure}
			\qquad
				\begin{subfigure}{0.01\textwidth}
	\captionsetup{labelformat=empty}
		\caption{${}$}
		\label{fig:mat_swr}
	\end{subfigure}
	\\
	\begin{subfigure}{\textwidth}
		\includegraphics[width=\textwidth]{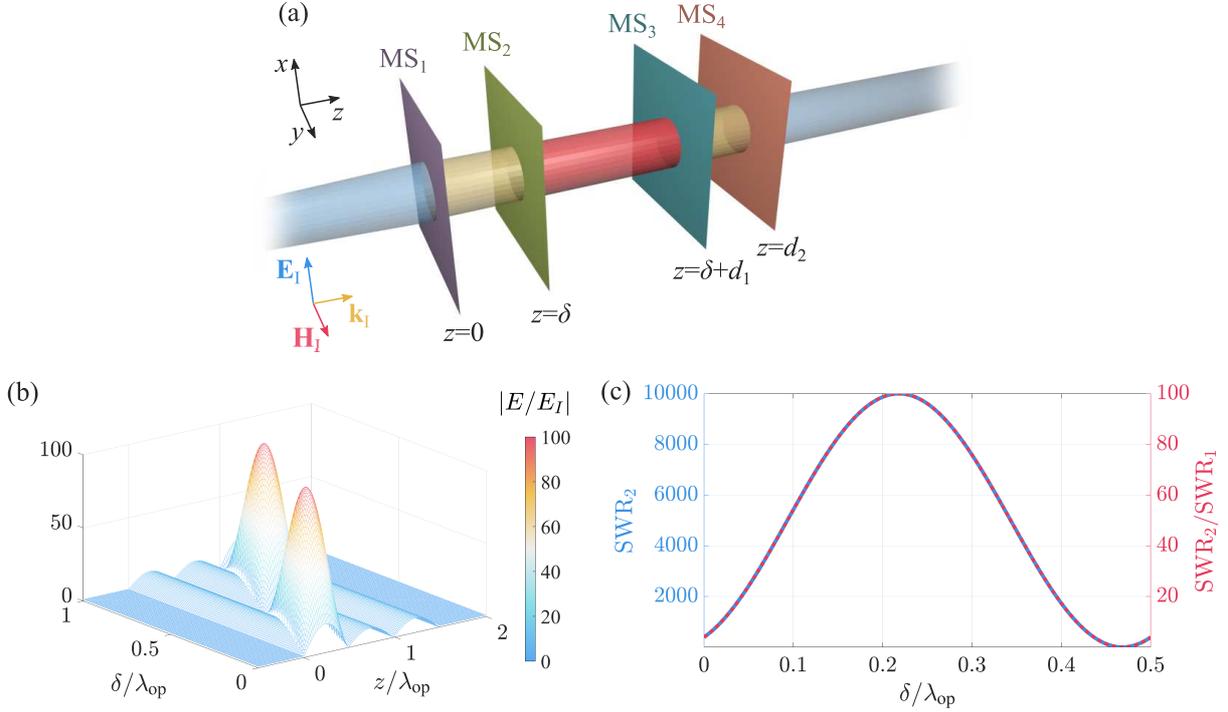}
	\end{subfigure}
	\caption{(a) Geometry of the problem. The cavity formed by four infinite metasurfaces is illuminated normally by a plane wave with electric field ${\bf E}_{\rm I}$. Due to the radiation by the currents induced in the metasurfaces, secondary fields are produced across the structure. (b) Distribution of the normalized total electric field across a matryoshka resonator with grid impedance of $Z\@{e}\approx j38.08\,\Omega$ as a function of distance  $\delta$. The outer resonator has a separation distance of $3\lambda\@{op}/2$, while the inner resonator metasurfaces are separated by $\lambda\@{op}/2$.
	(c) SWR of the inner resonator as a function of the separation between resonators $\delta$, using the reference grid impedance of $Z\@{e}\approx j38.08\,\Omega$. Notice that this is a periodic plot (with the period $\lambda\@{op}/2$).}
	\label{fig:mat_figure}
\end{figure}
In the case of conventional FP resonators, the SWR expression is similar to that in Equation~\r{s_e_swr} with the only difference  in the component $\vert 2 Z\@{e} - \eta\@{0} \vert$ which becomes $\vert 2 Z\@{e} + \eta\@{0} \vert$. It becomes  exactly the same expression in the case of pure-reactive metasurfaces for which $\Re (Z\@{e})=0$. 

The next step is designing a four-metasurface invisible resonator as shown in Fig.~\ref{fig:structure}.
The conventional approach for FP multi-mirror resonators requires finding the standing waves and the scattered fields from each metasurface pair. Unfortunately, due to the complexity of analytic expressions for each field (regardless of the method used), there is no straightforward approach to design a transparent FP compound resonator with a desired SWR~\cite{Stadt1985}. Even if the complexity of the analysis is reduced by assuming that the inner resonator is in the middle of the outer resonator (exploiting the symmetric profile of the standing wave inside the FP resonator), the the design approach of the   resonator is complex. 

We suggest an alternative approach for the four-metasurface scenario, where the inner metasurface pair is designed to be an invisible resonator, and therefore the standing wave of the outer resonator will not be disturbed by adding a second pair of metasurfaces. The scattering properties of such a resonator depend only on those of the outer resonator, which can be also designed to be an invisible resonator. One can add an {\it unlimited number} of outer metasurface pairs without compromising the zero-scattering condition of the entire structure. The advantage is that each new additional metasurface pair can be readily added to the existing ones without the need to recalculate the total scattering and do any optimizations. Such a compound structure can be referred to as an invisible ``matryoshka-doll'' resonator. We stress that the whole structure remains invisible, so that outside of the resonator the incident field is not perturbed at all. In constrast, a FP multi-mirror resonator can only achieve transparency (unitary transmission), while invisibility (zero-phase transmission) is achieved only for the trivial case when $\vert Z\@e \vert \rightarrow \infty$.
Due to the invisibility property, the inner resonator can be placed anywhere inside the outer resonator, i.e. at any distance $\delta$ measured from  the plane $z=0$. This extra degree of freedom allows us to modulate the fields across the structure, as shown in Fig.~\ref{fig:mat_fields}. Since this extra degree of freedom does not exist for FP multi-mirror resonators, they must be properly designed to achieve the desired field modulation without loosing transparency.

Similarly to a single resonator, the SWR of the invisible matryoshka resonator can be expressed in terms of the  sheet impedances:
\e {\rm SWR}\@{2} =\dfrac{\vert(2 Z\@{e}-\eta_0)^2 e^{-2j k_0 \delta}-\eta_0^2\vert+\vert\eta_0(2 Z\@{e}-\eta_0) e^{-2 j k_0 \delta}+2 Z\@{e}\eta_0+\eta_0^2\vert }{\vert(2 Z\@{e}-\eta_0)^2 e^{-2j k_0 \delta}-\eta_0^2\vert-\vert\eta_0(2 Z\@{e}-\eta_0) e^{-2 j k_0 \delta}+2 Z\@{e}\eta_0+\eta_0^2\vert }, \l{matroska_inner_SWR}\f
where  $k_0$ is the wavenumber of the wave in the medium. With the proper alignment,  ${\rm SWR}\@{2}$ can achieve the maximum of ${\rm SWR}\@{1}^2$, i.e. the product of the correspondent ratios for the inner and outer metasurface pairs. For example, using metasurfaces with grid reactance $\vert X\@{e}\vert \approx 38\,\Omega$ (which corresponds to  SWR of 100 in a single resonator configuration), the maximum value of ${\rm SWR}\@{2}=100^2$ is achieved for the separation  $\delta\approx 0.2185\lambda_{\rm op}$, as shown in Fig.~\ref{fig:mat_swr}.

\section{Conclusion}
In this work, we proposed an  intuitive approach for design of multi-metasurface resonators which remain invisible while the standing wave inside them can be tuned and dramatically enhanced. The structure remains non-scattering in both backward and forward directions regardless of the position of the inner resonator. 
Metasurfaces in this configuration can produce extreme standing wave ratios comparable to that of a single resonator whose mirrors possess very high reflectivities   (unavailable at optical frequencies).  In the talk, we will show the all-dielectric realization of the designed resonators in the visible range. The performance of the proposed structure suggests interesting possibilities for applications related with sensing, field localization, filtering, non-linear effects, and cloaking.



\end{document}